\documentclass[journal]{igtesymp}

\usepackage{graphicx}
\usepackage{amsmath}
\usepackage{siunitx}
\usepackage{xspace}
\usepackage[frozencache]{minted}
\usepackage{subcaption}
\usepackage{ninecolors}
\usepackage[nolist]{acronym}
\usepackage{tudacolors}
\usepackage{csquotes}
\usepackage{upgreek}
\usepackage{placeins}
\usepackage[markcase=noupper]{scrlayer-scrpage}
\setkomafont{pageheadfoot}{}

\usepackage{tikz}
\usepackage{pgfplots}
\usepackage{tikz-3dplot}
\usepackage{ifthen}
\usepackage{pgffor}
\usepackage{pgfmath}

\usetikzlibrary{decorations.pathmorphing, decorations.pathreplacing, decorations.shapes,decorations.markings,math,3d,circuits.ee.IEC,calc,arrows.meta,positioning, shapes,fit,calc,chains}
\pgfplotscreateplotcyclelist{myColorCycleList}{black,TUDa-2b,TUDa-9b,TUDa-3c,TUDa-7b,TUDa-11b,TUDa-10c,TUDa-8c}
\usetikzlibrary{pgfplots.colormaps,patterns, patterns.meta}
\usepgfplotslibrary{colorbrewer, patchplots}
\pgfplotscreateplotcyclelist{viridis few}{[samples of colormap=4 of viridis]}
\pgfplotscreateplotcyclelist{viridis many}{[samples of colormap=9 of viridis]}
\pgfplotsset{cycle list name={myColorCycleList}}

\pgfplotsset{compat=1.18}
\pgfplotsset{every axis plot/.style={thick,mark=none}}
\tikzset{>=Latex}
\pgfplotsset{field plot element/.style={line width=0pt,faceted color=none}}
\pgfplotsset{field plot node/.style={line width=0pt,faceted color=none,shader=interp}}
\usepackage[europeaninductors]{circuitikz}

\RequirePackage{fontawesome}
\newcommand{\zhref}[2]{\href{#1}{#2~\faExternalLink}}  %

\graphicspath{{.}}
\usepackage{csquotes}%
\usepackage[style=authoryear-ibid,backend=biber,firstinits=true, maxcitenames=2]{biblatex}
\newcommand{\cstems}{\textsc{CST EM STUDIO\textsuperscript{\textregistered}}\xspace}
\newcommand{\comsol}{\textsc{COMSOL Multiphysics\textsuperscript{\textregistered}}\xspace}

\usepackage[colorlinks=false, hidelinks]{hyperref}

\IEEEaftertitletext{\vspace{-1cm}\noindent%
	\begin{abstract}%
		\newline
		\textbf{Purpose} --- The purpose of this paper is to present the freely available finite element simulation software \emph{Pyrit}.\\
		\textbf{Design/methodology/approach} -- In a first step, the design principles and the objective of the software project are defined. Then, the software’s structure is established: The software is organized in packages for which an overview is given. The structure is based on the typical steps of a simulation workflow, i.e., problem definition, problem solving and post-processing. 
		State-of-the-art software engineering principles are applied to ensure a high code quality at all times. Finally, the modeling and simulation workflow of \emph{Pyrit} is demonstrated by three examples.\\
		\textbf{Findings} --- \emph{Pyrit} is a field simulation software based on the finite element method written in Python to solve coupled systems of partial differential equations. It is designed as a modular software that is easily modifiable and extendable. The framework can, therefore, be adapted to various activities, i.e., research, education and industry collaboration. \\
		\textbf{Research limitations/implications} --- The focus of \emph{Pyrit} are static and quasistatic electromagnetic problems as well as (coupled) heat conduction problems. It allows for both time domain as well as frequency domain simulations.\\
		\textbf{Originality/value} --- In research, problem specific modifications and direct access to the source code of simulation tools are essential. With \emph{Pyrit}, the authors present a computationally efficient and platform-independent simulation software for various electromagnetic and thermal field problems.
		\textbf{Paper type} --- Research paper
	\end{abstract}%
	\noindent%
	\vspace{\baselineskip}
}

\newcommand{\pyrit}{\emph{Pyrit}\xspace}

\newcommand{\getdp}{\emph{GetDP}\xspace}
\newcommand{\gmsh}{\emph{Gmsh}\xspace}
\let\module\textit
\let\function\texttt 
\let\class\texttt

\newminted{python}{fontsize={\footnotesize},breaklines,autogobble,breakafter=\],firstnumber=last,numbers=left,frame=leftline,xleftmargin=0.3cm,numbersep=0.1cm,tabsize=8,obeytabs=false}
\newmintinline{python}{}
\BeforeBeginEnvironment{pythoncode}{\vspace{0.2cm}}
\AfterEndEnvironment{pythoncode}{\vspace{0.2cm}}

\begin{document}

	\title{Pyrit: A Finite Element Based Field Simulation Software Written in Python\thanks{\textcopyright\ Jonas Bundschuh et. al. Published in COMPEL - The international journal for computation and mathematics in electrical and electronic engineering (DOI: \href{https://doi.org/10.1108/COMPEL-01-2023-0013}{10.1108/COMPEL-01-2023-0013}). Published by Emerald Publishing Limited.}}
	\author{Jonas~Bundschuh\IEEEauthorrefmark{1}\textsuperscript{,}\IEEEauthorrefmark{2}, M.\ Greta Ruppert\IEEEauthorrefmark{1}\textsuperscript{,}\IEEEauthorrefmark{2} and Yvonne Späck-Leigsnering\IEEEauthorrefmark{1}\textsuperscript{,}\IEEEauthorrefmark{2}\\
		\vspace{0.3cm} \normalsize{
			\IEEEauthorblockA{\IEEEauthorrefmark{1}Institut für Teilchenbeschleunigung und Elektromagnetische Felder (TEMF), TU Darmstadt, Germany\\
				\IEEEauthorrefmark{2}Graduate School of Excellence Computational Engineering, 64293 Darmstadt, Germany\\
				E-mail: \href{mailto:jonas.bundschuh@tu-darmstadt.de}{jonas.bundschuh@tu-darmstadt.de}}}% <-this % stops a space
	}
	
	\ihead{\footnotesize \textcopyright \usekomafont{pageheadfoot}{Jonas Bundschuh et. al. Published in COMPEL - The international journal for computation and mathematics in electrical\\and electronic engineering (DOI: \href{https://doi.org/10.1108/COMPEL-01-2023-0013}{10.1108/COMPEL-01-2023-0013}). Published by Emerald Publishing Limited.}}
	
	\maketitle
	
	\section{Introduction}\label{sec:introduction}
	
	Today, \ac{em} field simulation software is essential for the design of electric equipment, e.g., electric power converters and transmission systems, electronics and particle accelerator components.
	Many different tools for \ac{em} field simulation based on the \ac{fe} method have been developed over time. Commercial software tools, e.g., Ansys Maxwell\textsuperscript{\textregistered} \parencite{AnsysMaxwell}, \cstems \parencite{CSTEMS},  Flux2D/3D\textsuperscript{\textregistered} \parencite{Flux}, JMAG\textsuperscript{\textregistered} \parencite{Jmag}, and \comsol \parencite{Comsol}, are designed to solve universal \ac{em} field problems with standard computational methods.
	The design principles of commercial packages are, first of all, a broad applicability. Second, the software is made easy to use by powerful graphical user interfaces and template-based workflows. However, users cannot customize and access the internal routines and system solvers.
	
	In a research setting, however, problem-specific modifications and direct access to the \ac{fe} matrices is indispensable. 
	Various available freeware \ac{fe} tools offer different levels of abstraction: General \ac{fe} research tools with a focus on numerical mathematics are e.g., \textit{Deal} \parencite{Deal}, \mbox{\textit{FEniCS}} \parencite{Alnaes2015a}, and \getdp \parencite{Dular1998a}. However, these tools typically require an in-depth knowledge of the \ac{fe} method. Besides, research tools motivated by electrical engineering problems are, e.g., \textit{Agros} \parencite{Agros},  \textit{FEMM} \parencite{Meeker_2009aa},  \textit{openCFS} \parencite{openCFS}, and for electrical machine problems \textit{Pyleecan} \parencite{Bonneel_2018aa}. 
	These tools allow to study and further develop formulations, discretization techniques and solver strategies for electromagnetic field simulation, thereby covering a large fraction of research in a computational electromagnetics research group. Nevertheless, there is an additional need for a research code built from scratch, with access to all basic routines, with full flexibility and possibly adapted to a few specific lines of research.
	
	\zhref{https://git.rwth-aachen.de/jonas.bundschuh/pyrit-wiki}{\pyrit}\,\footnote{\pyrit's public wiki:\\ \href{https://git.rwth-aachen.de/jonas.bundschuh/pyrit-wiki}{https://git.rwth-aachen.de/jonas.bundschuh/pyrit-wiki}} is a \ac{fe} solver developed from the perspective of the electric field simulation workflow. In order to allow a fast prototyping and due to its popularity in industry and academia, it is written in Python\footnote{Python website: \href{https://www.python.org/}{https://www.python.org/}}. 
	In general, it solves static and quasistatic \ac{em} and heat transfer problems. 
	One of the central aims of \pyrit is to provide a user-friendly and computationally efficient \ac{fe} code with a template-based structure for standard and coupled \ac{em} and thermal problems. Furthermore, a plain user interface via Python scripts supports the implementation of new modeling and simulation ideas. 
	Its inherent structure supports students to learn how to implement and use a \ac{fe} solver.

	\section{Design principles}\label{sec:design_priciples}
	\pyrit is developed collaboratively based on the following principles: it is robust, portable, scriptable, extensible and thoroughly documented. Users organize the simulation workflow with Python scripts or Jupyter notebooks \parencite{Perkel2018a}. Therefore, the full capabilities of Python can be exploited, and studies can be tailored to the needs of individual users.
	
	The object-oriented structure of \pyrit allows, in contrast to other \ac{fe} software packages, different levels of abstraction: On one hand, the user can access and customize the basic and physics-independent \ac{fe} routines. On the other hand, \pyrit offers classes that serve as templates and predefined solvers for a set of selected \ac{em} and thermal problems.
	This allows to model and simulate field problems without detailed knowledge of the weak formulation or of the \ac{fe} method. 
	Additionally, \pyrit provides standard post-processing routines as well as the possibility to export results to Paraview and \LaTeX\ (for plotting with pgfplots). %
	\pyrit is user-friendly as all packages, modules, classes, functions and methods are thoroughly documented. The documentation of the whole software is stripped directly from the source files and is therefore always up-to-date. Supplementary information such as tutorials further extends the documentation.
	Furthermore, the geometry and mesh generation of the open source software \gmsh \parencite{Geuzaine2009a} is used. In order to make the geometry generation more convenient, the package \module{geometry} provides a unified interface to \gmsh.

	\section{Software structure}\label{sec:software_structure}
	\begin{figure}
		\centering
		\def\tmpwidth{2.45cm}
		\begin{tikzpicture}[node distance = 0.3cm]
			\begin{scope}[align=center,text width=1.8cm, draw=black,minimum width=\tmpwidth]
				\node[draw] (def) {Problem definition};
				\node[draw,right=of def] (solve) {Problem solving};
				\node[draw,right=of solve] (post) {Post-processing};
			\end{scope}
			
			\node at ($(def)!0.5!(solve)$) {$\Rightarrow$};
			\node at ($(solve)!0.5!(post)$) {$\Rightarrow$};
			
			\begin{scope}[minimum width=\tmpwidth]
				\node[draw,below=.3cm of def] (geometry) {geometry};
				\node[draw,below] (region) at (geometry.south) {region};
				\node[draw,below] (material) at (region.south) {material};
				\node[draw,below] (bdrycond) at (material.south) {bdrycond};
				\node[draw,below] (excitation) at (bdrycond.south) {excitation};
				\node[draw,below] at (excitation.south) {toolbox};
			\end{scope}
			\begin{scope}[minimum width=\tmpwidth]
				\node[draw,below] (problem) at ($(solve.north)!(region.north)!(solve.south)$){problem};
				\node[draw,below] (mesh) at (problem.south) {mesh};
				\node[draw,below] (shapefunction) at (mesh.south) {shapefunction};
				\node[draw,below] at (shapefunction.south) {toolbox};
			\end{scope}
			
			\begin{scope}[minimum width=\tmpwidth]
				\node[draw,below] (solution) at ($(post.north)!(mesh.north east)!(post.south)$) {solution};
				\node[draw,below] at (solution.south) {toolbox};
			\end{scope}
		\end{tikzpicture}
		\caption{Assignment of the packages to the steps \enquote{Problem definition}, \enquote{Problem solving} and \enquote{Post-processing} of a field simulation workflow.}
		\label{fig:module_assignment}
	\end{figure}
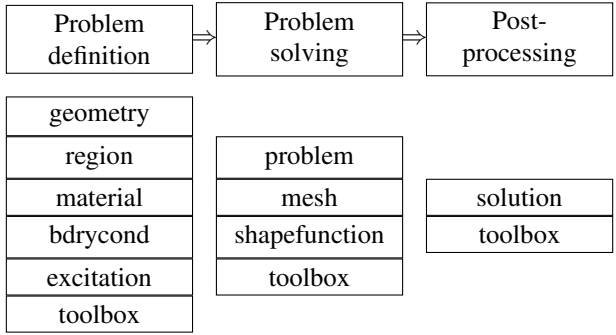
	\pyrit's modeling and simulation workflow is organized in packages. They are assigned to the three steps in Fig.~\ref{fig:module_assignment} that are described in this section. Since the \module{toolbox} package can be associated with all three steps, it is described here:
	\begin{LaTeXdescription}
		\item[toolbox] This package is a collection of different toolboxes with additional functionalities that are not solely associated to one package. It contains, inter alia, a material and geometry library (for the problem definition step), a circuit simulation and a time integration toolbox (for the problem solving step) and a post-processing and export toolbox (for the post-processing step).
	\end{LaTeXdescription} 
	
	\subsection{Problem definition}
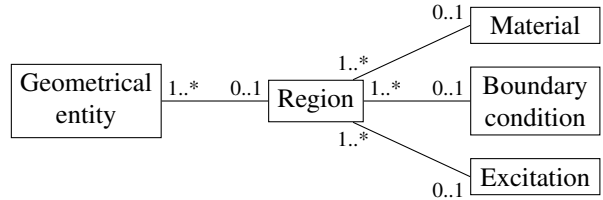
\begin{figure}
		\centering
		\def\mysize{\footnotesize}
		\def\acccolor{black}
		\def\widthright{1.7cm}
		\begin{tikzpicture}[node distance=1.4cm]
			\node[draw,align=center] (entity) {Geometrical\\entity};
			\node[draw,align=center,right=of entity] (region) {Region};
			\node[draw,align=center,right=of region,minimum width=\widthright] (bc) {Boundary\\condition};
			\node[draw,above right,minimum width=\widthright] (mat) at ($(bc.north west) + (0,0.3cm)$) {Material};
			\node[draw,below right,minimum width=\widthright] (exci) at ($(bc.south west) - (0,0.3cm)$) {Excitation};
			
			\draw (entity.east) -- (region.west) node[pos=0,above right=-0.07cm] {\mysize 1..*} node[above left=-0.07cm] {\mysize 0..1};
			\draw (region.30) -- (mat.west) node[pos=0,above] {\mysize 1..*}node[above left=-0.07cm] {\mysize 0..1};
			\draw (region.east) -- (bc.west) node[pos=0,above right=-0.07cm] {\mysize 1..*}node[above left=-0.07cm] {\mysize 0..1};
			\draw (region.330) -- (exci.west) node[pos=0,below] {\mysize 1..*}node[below left=-0.07cm] {\mysize 0..1};
			\tdplotsetmaincoords{75}{-40}
		\end{tikzpicture}
		\caption{UML class diagram of the assignment of materials, boundary conditions and excitations to physical groups. They, in turn, are then assigned to geometrical entities, as points, lines, triangles and tetrahedra. An object of a class can have at most one object of a class to its right but one or more objects of a class to its left.}
		\label{fig:regions}
	\end{figure}

	Before a simulation is executed, \pyrit defines a field problem. This first step is the choice of the problem type. 
	This includes the differential equation to be solved, the dimension of the problem and the time dependency. 
	Subsequently, the geometry is built and the materials, boundary conditions and excitations are assigned. The assignment relies on the concept of physical groups of \gmsh, as indicated by an UML class diagram in Fig.~\ref{fig:regions}. The pendant in \pyrit are regions. A region has one or more geometrical entities that share a material, a boundary condition or an excitation. 
	Consequently, a geometrical entity can have at most one region.
	
	The following packages are associated with the problem definition step:
	\begin{LaTeXdescription}
		\item[bdrycond] In this package, boundary conditions are defined. Besides the standard Dirichlet and Neumann boundary conditions, also Robin, periodic, anti-periodic and floating boundary conditions are supported \parencite{Gersem2004aa}. 
		There is a container class that manages all boundary conditions of a problem and can apply them to the system of equations of the corresponding problem.
		\item[excitation] In this package, excitations for problems are defined. Its structure is similar to \module{bdrycond}. Typical excitations are charges, charge densities, currents or current densities. In addition to that, the field-circuit coupling models solid and stranded conductor \parencite{Gersem2004ab} are also implemented in this package. 
		Like in the previous package, there is also a container class that manages all excitations of a problem.
		\item[geometry] In this package, the interaction between \pyrit and \gmsh is handled via the \gmsh Python API. This includes primary the generation of geometries and meshes and the extraction of the relevant data for \pyrit. There are also several wrapper classes for geometrical entities to simplify the geometry generation.
		\item[material] In this package, the management of material information is implemented. Materials can be defined by giving them a set of material properties, e.g., a conductivity or a permeability. The values of the material properties can be constants or functions that can depend, in particular, on space, time or other field values. They can be scalar or tensor valued. With this, a nonlinear, inhomogeneous and anisotropic material can be defined.
		A container class manages all the materials of a problem.
		\item[region] In this package, classes for the organization of materials, boundary conditions and excitations on the mesh are provided. Regions can be seen as discrete counterpart to physical groups and are defined on the mesh. The package includes a class for defining a region and a container class that manages different regions.
	\end{LaTeXdescription}
	
	\subsection{Problem solving}
	Now that the problem is completely defined, it can be solved. 
	For that, a mesh object and a shape function object are required. 
	The former can be generated and extracted from \gmsh with the \module{geometry} package. 
	The latter is fixed by the chosen problem. 
	With the shape function object, the required matrices and vectors can be computed and assembled into the final system of equations.
	This can be solved using a predefined \function{solve} routine defined in the problem class or any other solve method. Currently, there exist wrappers for direct and iterative solvers for sparse matrices from \zhref{https://scipy.org/}{SciPy}\,\footnote{\href{https://docs.scipy.org/doc/scipy/reference/sparse.linalg.html}{https://docs.scipy.org/doc/scipy/reference/sparse.linalg.html}} and \zhref{https://www.intel.com/content/www/us/en/docs/onemkl/developer-reference-fortran/2023-0/onemkl-pardiso-parallel-direct-sparse-solver-iface.html}{Pardiso}\,\footnote{\href{https://www.intel.com/content/www/us/en/docs/onemkl/developer-reference-fortran/2023-0/onemkl-pardiso-parallel-direct-sparse-solver-iface.html}{https://www.intel.com/content/www/us/en/docs/onemkl/developer-reference-fortran/2023-0/onemkl-pardiso-parallel-direct-sparse-solver-iface.html}}.
	
	The relevant packages in this step are:
	\begin{LaTeXdescription}
		\item[mesh] In this package, classes for different kinds of meshes are defined, e.g., a triangular mesh for two-dimensional domains or a tetrahedral mesh for three-dimensional domains. These classes are responsible for the mesh data. They store the coordinates of all nodes and the definition of the higher dimensional entities (edges, triangles and tetrahedra).
		\item[problem] This package collects classes that represent standard problem formulations. They provide a framework for organizing all the data needed for defining a problem. Furthermore, the classes in the \module{problem} package implement a \function{solve} method that  provides a convenient way to solve a problem, i.e., to setup and solve the system of equations. 
		There are separate classes for static, harmonic and transient problems per problem type, i.e., the underlying differential equation.
		\item[shapefunction] This package provides the core of the \ac{fe} method, i.e., the different shape functions. There is one class per \ac{fe} shape function and dimension of the domain. For example, there is a class for nodal shape functions in cylindrical coordinates or a class for edge shape functions in two dimensional Cartesian coordinates. These classes implement the routines to compute the \ac{fe} matrices. 
		Currently, nodal shape functions are implemented for two and three dimensional Cartesian and cylindrical coordinates. Furthermore, edge functions are implemented for two dimensional Cartesian and cylindrical coordinates.
	\end{LaTeXdescription}
	
	\subsection{Post-processing}\label{ssec:post_processing}
	The result from the problem solving step is a vector or several vectors, depending on the problem, respectively. 
	Each problem class is equipped with a corresponding solution class. 
	These implement standard post-processing routines and allow for a straight-forward visualization of the results. This step also includes the export of the data to files of different formats. 
	
	The post-processing step comprises the following package:
	\begin{LaTeXdescription}
		\item[solution] This package collects solution classes for standard problem formulations. 
		For each class in the \module{problem} package, there is an associated class in this package. 
		The problems \function{solve} routine returns a solution object of correct type that contains the simulation result.
		In the solution class, standard post-processing and visualization routines, specific for this problem type, are implemented. 
		In the case of an electrostatic problem this includes for example the computation and visualization of the electric displacement field from the electric scalar potential.
	\end{LaTeXdescription}
	
	\section{Quality management}

	In the development of \pyrit, state-of-the-art software engineering principles were applied to ensure a good code quality and a high usability and reproducibility.
	This section addresses testing, linting, and the documentation of the code. \Ac{CI} ensures that the code fulfills these requirements.
	
	A dedicated focus lies on developing the software collaboratively. 
	For that reason, the version control system Git\footnote{Git website: \href{https://git-scm.com/}{https://git-scm.com/}} is used. It keeps track of the evolving software project and serves as a backup. It also helps distributing the code between the developers and users. 
	
	In order to ensure the correctness and robustness of the code in presence and in future, there are many tests for \pyrit. They range from testing the functionality of single functions and methods over the interaction of modules to integration tests, i.e., the test of whole simulation procedures. 
	The tests of the shape functions cover, in particular, the consistency and convergence of the implemented \ac{fe} method. 
	
	The coding style, which can vary depending on the collaborator, and the readability of the code are optimized through the use of a linter, i.e., a static code analyzer that finds discrepancies between the code and a predefined style. 
	
	A crucial part for every software is its documentation, since it guides the user through the software's functionalities. 
	In \pyrit, the documentation is automatically generated based on the provided docstrings of the source files. This implies that the documentation file is always up-to-date. Furthermore, the documentation is enriched by application examples and tutorials of \ac{em} and thermal field problems.
	
	The tasks in the described aspects of quality management, i.e., testing, linting and documentation, have to be executed regularly, but at least after a commit has been pushed. This is automated on Git with \ac{CI}.  After new commits have been pushed, a so called pipeline is started where first \pyrit is installed with the recent dependencies and, then, tested and linted. The single tasks are executed via Docker\footnote{Docker website: \href{https://www.docker.com/}{https://www.docker.com/}} in containers on a separate machine \parencite{merkel2014docker}.
	
	\section{Examples}
	Three examples show the modeling and simulation workflow in \pyrit. The first two examples use a predefined problem class of \pyrit, i.e., the user models the \ac{em} problem. Subsequently, the predefined template is used to setup, solve and post-process the \ac{fe} system. The third example shows how \pyrit can be customized. There, a system of equations from the \ac{fe} formulation is built and extended by a field-circuit coupling. 
	
	\subsection{Template-based simulation example}
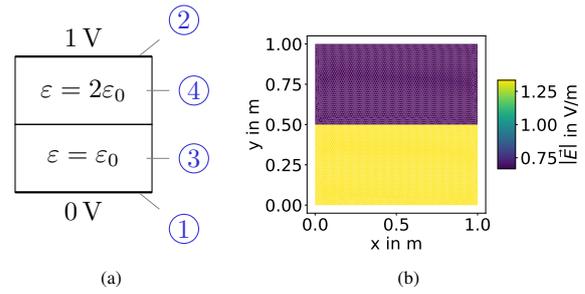
\begin{figure}
		\centering
		\subcaptionbox{\label{fig:place_cap}}{
			\begin{tikzpicture}[x=1.8cm,y=1.8cm]
				\draw (0,0) rectangle (1,0.5) node[pos=0.5] {$\varepsilon=\varepsilon_0$};
				\draw (0,0.5) rectangle (1,1) node[pos=0.5] {$\varepsilon=2\varepsilon_0$};
				\draw[thick] (0,0) -- (1,0) node[pos=0.5,below] {\SI{0}{\volt}};
				\draw[thick] (0,1) -- (1,1) node[pos=0.5,above] {\SI{1}{\volt}};
				\begin{scope}[blue3,pin distance=.3cm]
					\node[inner sep=0,outer sep=0,pin=330:\tikz{\node[draw,inner sep=1pt,circle] {$1$};}] at (0.9,0) {};
					\node[inner sep=0,outer sep=0,pin=30:\tikz{\node[draw,inner sep=1pt,circle] {$2$};}] at (0.9,1) {};
					\node[inner sep=0,outer sep=0,pin=0:\tikz{\node[draw,inner sep=1pt,circle] {$3$};}] at (0.95,0.25) {};
					\node[inner sep=0,outer sep=0,pin=0:\tikz{\node[draw,inner sep=1pt,circle] {$4$};}] at (0.95,0.75) {};
				\end{scope}
			\end{tikzpicture}
		}
		\subcaptionbox{\label{fig:e_field}}{
			\includegraphics[width=4.4cm]{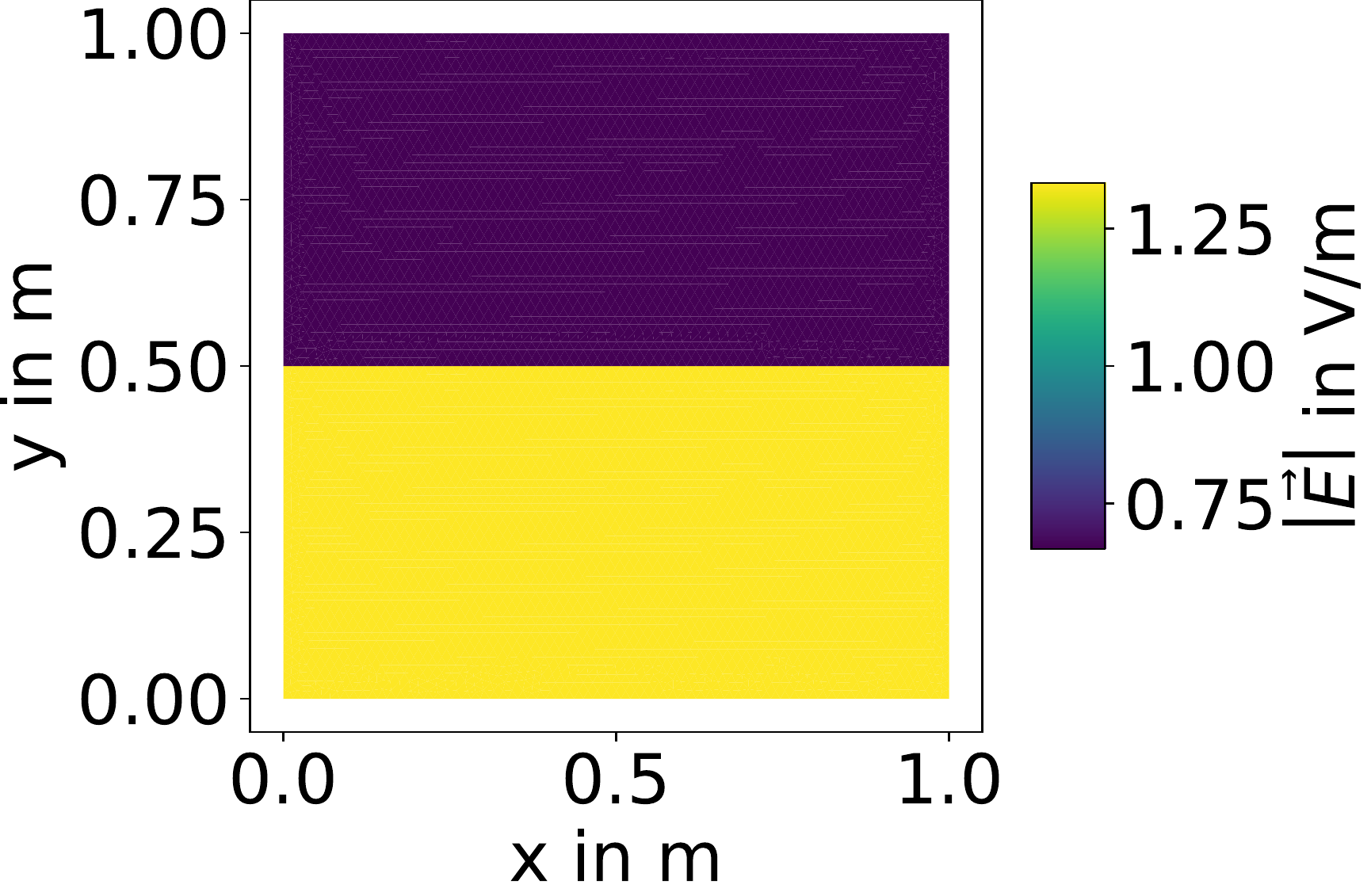}
		}
		\caption{Plate capacitor with two materials. In (\subref{fig:place_cap}) the model is shown with the region IDs in blue. In (\subref{fig:e_field}) the electric field strength is plotted.}
		\label{fig:fig_frist_example}
	\end{figure}

	As a starting point, a simple example shows the basic procedure of the problem definition.
	It is an electrostatic simulation of a plate capacitor in 2D Cartesian coordinates filled with two materials (see Fig.~\ref{fig:place_cap}). The bottom plate is grounded and the top plate is set to a potential of \SI{1}{\volt}. On the left and right side homogeneous Neumann boundary conditions are imposed. 
	
	We start at a point where the mesh and the regions already exist. The geometry was build with the \gmsh Python API and imported into \pyrit with the \module{geometry} package.
	\begin{pythoncode}
		model = PlateCapacitor()
		mesh, regions = model.create_geometry()
	\end{pythoncode}
	
	Next, the materials and boundary conditions of this example are defined:
	\begin{pythoncode}
		materials = Materials(
		material_bot := Mat("Mat_bot", Permittivity(eps_0)),
		material_top := Mat("Mat_top",  Permittivity(2 * eps_0)))
		
		boundary_conditions = BdryCond(
		bc_ground := BCDirichlet(0),
		bc_voltage := BCDirichlet(1))
	\end{pythoncode}
	
	With the region IDs from Fig.~\ref{fig:place_cap}, materials and boundary conditions can be assigned to the regions:
	\begin{pythoncode}
		regions.get_regi(1).bc = bc_ground.ID
		regions.get_regi(2).bc = bc_voltage.ID
		regions.get_regi(3).mat = material_bot.ID
		regions.get_regi(4).mat = material_top.ID
	\end{pythoncode}
	
	With that, the problem is completely defined and all data structures necessary to instantiate an object of the problem class \class{Electric\-Problem\-Cart\-Static} are available. The \function{solve} routine then solves the problem and returns an object of the associate solution class \class{Electric\-Solution\-Cart\-Static}:
	\begin{pythoncode}
		problem = ElectricProblemCartStatic("Plate Capacitor", mesh, regions, materials, boundary_conditions, Excitations())
		solution: = problem.solve()
	\end{pythoncode}
	
	The solution object can be used to plot for instance the electric field strength (see Fig.~\ref{fig:e_field}):
	\begin{pythoncode}
		solution.plot_e_field(plot_type='abs')
	\end{pythoncode}
	
	\setcounter{FancyVerbLine}{0}%
	\subsection{Template-based nonlinear simulation example}
	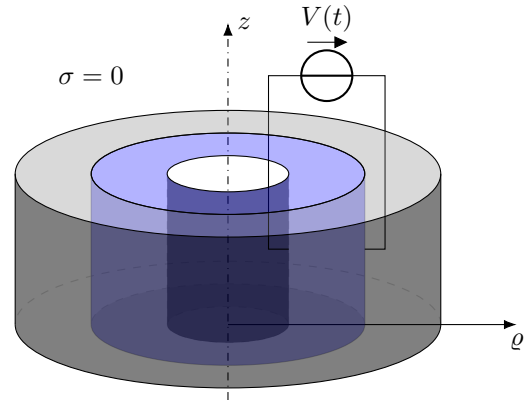
\begin{figure}
		\centering
		\def\outerradius{2.8}
		\def\radiusi{0.8}
		\def\radiuso{1.8}
		\def\height{2}
		\tikzset{behind/.style={draw=gray!30,dashed},
			intermediate/.style={draw=gray!50}}
		\def\intermediate{gray!50}
		\pgfdeclarelayer{behind}\pgfdeclarelayer{intermediate}
		\pgfsetlayers{behind,intermediate,main}
		\tikzset{level/.style={%
				execute at begin scope={\pgfonlayer{#1}},
				execute at end scope={\endpgfonlayer}
		}}
		\ctikzset{sources/scale=0.8}%
		\begin{tikzpicture}[z={(0,0.3)},circuit ee IEC]%
			\pgfmathparse{0.5*\height}
			\pgfmathsetmacro{\halfheight}{\pgfmathresult}
			
			\pgfmathparse{0.5*(\radiusi+\radiuso)}
			\pgfmathsetmacro{\halfradius}{\pgfmathresult}
			
			\draw[->,dashdotted] (0,-1,0) -- ++(0,\height+3,0) node[pos=1,right] {$z$};
			\draw[->] (0,0,0) -- ++(\outerradius+1,0,0) node[pos=1,below] {$\varrho$};
			
			\fill[black,opacity=0.5] plot[variable=\x,domain=180:360,smooth] ({\radiusi*cos(\x)},0,{\radiusi*sin(\x)}) -- plot[variable=\x,domain=0:-180,smooth] ({\radiusi*cos(\x)},\height,{\radiusi*sin(\x)});
			
			\fill[blue,opacity=0.3] plot[variable=\x,domain=180:360,smooth] ({\radiuso*cos(\x)},0,{\radiuso*sin(\x)}) -- plot[variable=\x,domain=0:-180,smooth] ({\radiuso*cos(\x)},\height,{\radiuso*sin(\x)});
			
			\fill[black,opacity=0.5] plot[variable=\x,domain=180:360,smooth] ({\outerradius*cos(\x)},0,{\outerradius*sin(\x)}) -- plot[variable=\x,domain=0:-180,smooth] ({\outerradius*cos(\x)},\height,{\outerradius*sin(\x)});
			
			to [voltage source={direction info={<-,volt}}]
			\draw (\radiusi,\height/2) -- ++(-\radiusi/3,0,0) -- ++(0,2.3,0) to[european voltage source, v^={$V(t)$}] (\radiuso+\radiusi/3,\height+1.3,0) -- ++(0,-2.3,0) -- (\radiuso,\height/2);

			\node at (-\radiuso,\height+1.3,0) {$\sigma=0$};
			
			\begin{scope}[canvas is xz plane at y=\height]
				\filldraw[fill=blue,fill opacity=.3,even odd rule] (0,0) circle (\radiuso cm) (0,0) circle (\radiusi cm);
				\filldraw[fill=gray,fill opacity=.3,even odd rule] (0,0) circle (\radiuso cm) (0,0) circle (\outerradius cm);
			\end{scope}
			\begin{scope}[canvas is xz plane at y=0]
				\begin{scope}[level=intermediate]
					\draw[\intermediate] (-\radiuso,0) arc [start angle=180, end angle=360, radius=\radiuso cm];
				\end{scope}
				\begin{scope}[level=behind]
					\draw[behind] (\outerradius,0) arc [start angle=0, end angle=180, radius=\outerradius cm];
					\draw[behind] (-\radiusi,0) arc [start angle=180, end angle=360, radius=\radiusi cm];
					
					\draw[behind] (\radiusi,0) arc [start angle=0, end angle=180, radius=\radiusi cm];
					\draw[behind] (\radiuso,0) arc [start angle=0, end angle=180, radius=\radiuso cm];
					\fill[fill=gray!30,opacity=.5,even odd rule] (0,0) circle (\radiuso cm) (0,0) circle (\radiusi cm);
				\end{scope}
				
				\draw (-\outerradius,0) arc [start angle=180, end angle=360, radius=\outerradius cm];
			\end{scope}
			\begin{scope}[level=intermediate,\intermediate]
				\draw[\intermediate] (\radiuso,0,0) -- +(0,\height,0) (-\radiuso,0,0) -- +(0,\height,0);
			\end{scope}
			\begin{scope}[level=behind]
				\draw[behind] (\radiusi,0,0) -- +(0,\height,0) (-\radiusi,0,0) -- +(0,\height,0);
				
			\end{scope}
			
			\draw (\outerradius,0,0) -- +(0,\height,0) (-\outerradius,0,0) -- +(0,\height,0);
		\end{tikzpicture}
		\caption{Schematic of the second example. The cylindrical resistor consists of a nonlinear material (blue) with an inner radius of \SI{0.1}{\meter} and an outer radius of \SI{0.3}{\meter}. The material is located between two electrodes to which a transient voltage $V(t)$ is applied, where $t$ denotes the time. The resistor is surrounded by a layer of soil (gray) with a thickness of \SI{0.7}{\meter}. The inner surface of the resistor and the outer surface of the soil are at a fixed temperature of \SI{60}{\degreeCelsius} and \SI{20}{\degreeCelsius}, respectively. The surrounding space is non-conductive.}
		\label{fig:ex1_setup}
	\end{figure}
	
	\begin{figure}
		\centering
		\newlength\fwidth
		\setlength{\fwidth}{0.6 \linewidth}
		\definecolor{mycolor1}{rgb}{0.24220,0.15040,0.66030}%
		\definecolor{mycolor2}{rgb}{0.91840,0.73080,0.18900}%
		\definecolor{mycolor3}{rgb}{0.10850,0.66690,0.87340}%
		\begin{tikzpicture}
			
			\begin{axis}[%
				width=1.15\fwidth,
				height=0.75\fwidth,
				at={(0\fwidth,0\fwidth)},
				scale only axis,
				xmin=0,
				xmax=2,
				xlabel style={font=\color{white!15!black}},
				xlabel={electric field strength $E$ in \unit{\kilo\volt\per\milli\meter}},
				ymode=log,
				ymin=8e-11,
				ymax=1e-3,
				yminorticks=true,
				ylabel style={font=\color{white!15!black}},
				ylabel={conductivity $\sigma$ in \unit{\siemens\per\meter}},
				axis background/.style={fill=white},
				xmajorgrids,
				ymajorgrids,
				yminorgrids,
				legend style={legend cell align=left, align=left, draw=white!15!black, legend pos = north west}
				]
				\addplot [color=mycolor1, line width=2.0pt]
				table[row sep=crcr, x expr = \thisrowno{0} * 1e-6]{%
					0	1.00053648068057e-10\\
					30303.0303030303	1.00074324471879e-10\\
					60606.0606060606	1.00102969730693e-10\\
					90909.0909090909	1.00142655106334e-10\\
					121212.121212121	1.0019763555004e-10\\
					151515.151515152	1.00273805906031e-10\\
					181818.181818182	1.00379332939655e-10\\
					212121.212121212	1.00525530954362e-10\\
					242424.242424242	1.00728074878608e-10\\
					272727.272727273	1.01008680886359e-10\\
					303030.303030303	1.01397434742481e-10\\
					333333.333333333	1.01936017511452e-10\\
					363636.363636364	1.02682174480582e-10\\
					393939.393939394	1.03715906442733e-10\\
					424242.424242424	1.05148047149983e-10\\
					454545.454545455	1.07132146587977e-10\\
					484848.484848485	1.09880934161241e-10\\
					515151.515151515	1.13689126924503e-10\\
					545454.545454545	1.18965028292447e-10\\
					575757.575757576	1.26274305149882e-10\\
					606060.606060606	1.3640063701682e-10\\
					636363.636363636	1.50429739895365e-10\\
					666666.666666667	1.69865773589879e-10\\
					696969.696969697	1.96792613352084e-10\\
					727272.727272727	2.34097276921649e-10\\
					757575.757575758	2.85779462146335e-10\\
					787878.787878788	3.5738038287735e-10\\
					818181.818181818	4.56576881621245e-10\\
					848484.848484849	5.94004517867331e-10\\
					878787.878787879	7.8439788100742e-10\\
					909090.909090909	1.0481703884532e-09\\
					939393.939393939	1.41360294947594e-09\\
					969696.96969697	1.91987615532311e-09\\
					1000000	2.62127109509402e-09\\
					1030303.03030303	3.59298919061209e-09\\
					1060606.06060606	4.93921502446369e-09\\
					1090909.09090909	6.80428661186021e-09\\
					1121212.12121212	9.38817071747865e-09\\
					1151515.15151515	1.29679023532892e-08\\
					1181818.18181818	1.79272869795703e-08\\
					1212121.21212121	2.47980496884915e-08\\
					1242424.24242424	3.43168426818642e-08\\
					1272727.27272727	4.7504222097393e-08\\
					1303030.3030303	6.57740596957454e-08\\
					1333333.33333333	9.1085116077733e-08\\
					1363636.36363636	1.26151018736528e-07\\
					1393939.39393939	1.74731142950922e-07\\
					1424242.42424242	2.4203355369087e-07\\
					1454545.45454545	3.35273154689478e-07\\
					1484848.48484848	4.64444780124319e-07\\
					1515151.51515152	6.43393910905357e-07\\
					1545454.54545455	8.91299418109337e-07\\
					1575757.57575758	1.23472654810059e-06\\
					1606060.60606061	1.71046880458131e-06\\
					1636363.63636364	2.3694806254977e-06\\
					1666666.66666667	3.28231713421389e-06\\
					1696969.6969697	4.54665390203097e-06\\
					1727272.72727273	6.29767323485685e-06\\
					1757575.75757576	8.72239277352653e-06\\
					1787878.78787879	1.20794003409468e-05\\
					1818181.81818182	1.67259726642688e-05\\
					1848484.84848485	2.31552216205591e-05\\
					1878787.87878788	3.20467480219687e-05\\
					1909090.90909091	4.43352793368453e-05\\
					1939393.93939394	6.13028454500743e-05\\
					1969696.96969697	8.47009817882759e-05\\
					2000000	0.000116909733797679\\
					2030303.03030303	0.000161138842995848\\
					2060606.06060606	0.000221671514268356\\
					2090909.09090909	0.000304139349099001\\
					2121212.12121212	0.000415793455560373\\
					2151515.15151515	0.000565695301501934\\
					2181818.18181818	0.000764687236809748\\
					2212121.21212121	0.00102492256069253\\
					2242424.24242424	0.00135867074244011\\
					2272727.27272727	0.00177614354850854\\
					2303030.3030303	0.0022823362271732\\
					2333333.33333333	0.00287343736614203\\
					2363636.36363636	0.00353410659302064\\
					2393939.39393939	0.00423733693721845\\
					2424242.42424242	0.00494801177869064\\
					2454545.45454545	0.00562952083406273\\
					2484848.48484848	0.00625097789376391\\
					2515151.51515152	0.00679219578871147\\
					2545454.54545455	0.00724497150071101\\
					2575757.57575758	0.00761119661452739\\
					2606060.60606061	0.00789941978432518\\
					2636363.63636364	0.00812140779673517\\
					2666666.66666667	0.00828955427041279\\
					2696969.6969697	0.0084153161846427\\
					2727272.72727273	0.00850848974583747\\
					2757575.75757576	0.00857703568465424\\
					2787878.78787879	0.00862720307676784\\
					2818181.81818182	0.00866378057585047\\
					2848484.84848485	0.00869037587028906\\
					2878787.87878788	0.00870967428021159\\
					2909090.90909091	0.00872365739177105\\
					2939393.93939394	0.00873377845959663\\
					2969696.96969697	0.00874109854260011\\
					3000000	0.00874638987349994\\
				};
				\addlegendentry{$T = \SI{20}{\degreeCelsius}$}
				
				\addplot [color=mycolor2, line width=2.0pt]
				table[row sep=crcr, x expr = \thisrowno{0} * 1e-6]{%
					0	2.24700426506627e-10\\
					30303.0303030303	2.2474686156269e-10\\
					60606.0606060606	2.24811193068811e-10\\
					90909.0909090909	2.24900318463083e-10\\
					121212.121212121	2.25023793512608e-10\\
					151515.151515152	2.25194856855212e-10\\
					181818.181818182	2.25431849407919e-10\\
					212121.212121212	2.25760180827048e-10\\
					242424.242424242	2.26215053858098e-10\\
					272727.272727273	2.26845238672341e-10\\
					303030.303030303	2.27718301863571e-10\\
					333333.333333333	2.28927850742935e-10\\
					363636.363636364	2.30603569644163e-10\\
					393939.393939394	2.32925124302828e-10\\
					424242.424242424	2.36141425096955e-10\\
					454545.454545455	2.40597314497864e-10\\
					484848.484848485	2.46770540082487e-10\\
					515151.515151515	2.5532297724642e-10\\
					545454.545454545	2.67171593568512e-10\\
					575757.575757576	2.83586763420224e-10\\
					606060.606060606	3.06328473826435e-10\\
					636363.636363636	3.37835025173471e-10\\
					666666.666666667	3.81484458703208e-10\\
					696969.696969697	4.41956740282862e-10\\
					727272.727272727	5.25735532726948e-10\\
					757575.757575758	6.41803355210365e-10\\
					787878.787878788	8.0260466268076e-10\\
					818181.818181818	1.02538010371771e-09\\
					848484.848484849	1.33401501183509e-09\\
					878787.878787879	1.76160031959427e-09\\
					909090.909090909	2.35398046832681e-09\\
					939393.939393939	3.17466870815356e-09\\
					969696.96969697	4.31165678884162e-09\\
					1000000	5.88684914973295e-09\\
					1030303.03030303	8.06913310162441e-09\\
					1060606.06060606	1.10924863214384e-08\\
					1090909.09090909	1.52810630424824e-08\\
					1121212.12121212	2.10839485122979e-08\\
					1151515.15151515	2.91233078048123e-08\\
					1181818.18181818	4.02610910066577e-08\\
					1212121.21212121	5.56914460304974e-08\\
					1242424.24242424	7.70687459764651e-08\\
					1272727.27272727	1.06684955244102e-07\\
					1303030.3030303	1.47715346237583e-07\\
					1333333.33333333	2.0455890241154e-07\\
					1363636.36363636	2.83309886862517e-07\\
					1393939.39393939	3.92411102475306e-07\\
					1424242.42424242	5.43558818627585e-07\\
					1454545.45454545	7.52956262061572e-07\\
					1484848.48484848	1.04304982574673e-06\\
					1515151.51515152	1.44493368291642e-06\\
					1545454.54545455	2.00167973143816e-06\\
					1575757.57575758	2.77294818664222e-06\\
					1606060.60606061	3.84136987843032e-06\\
					1636363.63636364	5.32137825485749e-06\\
					1666666.66666667	7.37142597225636e-06\\
					1696969.6969697	1.0210872773669e-05\\
					1727272.72727273	1.41433110056033e-05\\
					1757575.75757576	1.9588744780566e-05\\
					1787878.78787879	2.71279104856706e-05\\
					1818181.81818182	3.75631799936266e-05\\
					1848484.84848485	5.20019836803551e-05\\
					1878787.87878788	7.19705686672083e-05\\
					1909090.90909091	9.95681453764015e-05\\
					1939393.93939394	0.000137673895801699\\
					1969696.96969697	0.000190221417218188\\
					2000000	0.000262555814349178\\
					2030303.03030303	0.000361885522887908\\
					2060606.06060606	0.000497829761955194\\
					2090909.09090909	0.000683035978993071\\
					2121212.12121212	0.000933788708428996\\
					2151515.15151515	0.00127043818965819\\
					2181818.18181818	0.00171733416595113\\
					2212121.21212121	0.00230177050982914\\
					2242424.24242424	0.00305130198851627\\
					2272727.27272727	0.00398886218137087\\
					2303030.3030303	0.00512566941415776\\
					2333333.33333333	0.00645316401929703\\
					2363636.36363636	0.00793689459710115\\
					2393939.39393939	0.00951620890822119\\
					2424242.42424242	0.0111122420671292\\
					2454545.45454545	0.0126427747200323\\
					2484848.48484848	0.0140384426348636\\
					2515151.51515152	0.0152539094786614\\
					2545454.54545455	0.0162707529177825\\
					2575757.57575758	0.0170932210722271\\
					2606060.60606061	0.0177405125047054\\
					2636363.63636364	0.0182390530580089\\
					2666666.66666667	0.018616676313937\\
					2696969.6969697	0.0188991123501176\\
					2727272.72727273	0.0191083614813973\\
					2757575.75757576	0.0192623019122045\\
					2787878.78787879	0.0193749678131718\\
					2818181.81818182	0.0194571135400205\\
					2848484.84848485	0.019516841237299\\
					2878787.87878788	0.0195601815954391\\
					2909090.90909091	0.0195915848595076\\
					2939393.93939394	0.0196143147479327\\
					2969696.96969697	0.0196307541862211\\
					3000000	0.0196426374541772\\
				};
				\addlegendentry{$T = \SI{40}{\degreeCelsius}$}
				
				\addplot [color=mycolor3, line width=2.0pt]
				table[row sep=crcr, x expr = \thisrowno{0} * 1e-6]{%
					0	4.57917399681911e-10\\
					30303.0303030303	4.58012029765428e-10\\
					60606.0606060606	4.58143131056415e-10\\
					90909.0909090909	4.5832475985626e-10\\
					121212.121212121	4.58576389879768e-10\\
					151515.151515152	4.58925000170562e-10\\
					181818.181818182	4.59407967716136e-10\\
					212121.212121212	4.60077074900394e-10\\
					242424.242424242	4.61004061461133e-10\\
					272727.272727273	4.62288316217314e-10\\
					303030.303030303	4.64067533250863e-10\\
					333333.333333333	4.66532475068008e-10\\
					363636.363636364	4.69947425603606e-10\\
					393939.393939394	4.74678525980419e-10\\
					424242.424242424	4.81233031101477e-10\\
					454545.454545455	4.90313696053726e-10\\
					484848.484848485	5.02894123475732e-10\\
					515151.515151515	5.203231504159e-10\\
					545454.545454545	5.44469466737561e-10\\
					575757.575757576	5.77921970636618e-10\\
					606060.606060606	6.2426734280806e-10\\
					636363.636363636	6.88474599955183e-10\\
					666666.666666667	7.77427858345799e-10\\
					696969.696969697	9.00664428762237e-10\\
					727272.727272727	1.07139737921062e-09\\
					757575.757575758	1.30793220152784e-09\\
					787878.787878788	1.63562947263259e-09\\
					818181.818181818	2.08962394099477e-09\\
					848484.848484849	2.71859157035531e-09\\
					878787.878787879	3.58996843116202e-09\\
					909090.909090909	4.79718099211722e-09\\
					939393.939393939	6.46966301884763e-09\\
					969696.96969697	8.78673305503933e-09\\
					1000000	1.19968203749087e-08\\
					1030303.03030303	1.6444100730152e-08\\
					1060606.06060606	2.2605397645609e-08\\
					1090909.09090909	3.11413056111067e-08\\
					1121212.12121212	4.29670162530555e-08\\
					1151515.15151515	5.93504408845538e-08\\
					1181818.18181818	8.2048149123481e-08\\
					1212121.21212121	1.13493697129492e-07\\
					1242424.24242424	1.57058534792091e-07\\
					1272727.27272727	2.17413460446277e-07\\
					1303030.3030303	3.01029367384101e-07\\
					1333333.33333333	4.1687095182846e-07\\
					1363636.36363636	5.77357723406165e-07\\
					1393939.39393939	7.99695285164512e-07\\
					1424242.42424242	1.10771948531559e-06\\
					1454545.45454545	1.53445090851787e-06\\
					1484848.48484848	2.12563309901206e-06\\
					1515151.51515152	2.94463292785241e-06\\
					1545454.54545455	4.07922669247412e-06\\
					1575757.57575758	5.65099605203656e-06\\
					1606060.60606061	7.82833452207691e-06\\
					1636363.63636364	1.08444462303517e-05\\
					1666666.66666667	1.50222421276258e-05\\
					1696969.6969697	2.08087558252295e-05\\
					1727272.72727273	2.88226787072316e-05\\
					1757575.75757576	3.99199379031212e-05\\
					1787878.78787879	5.52840171312963e-05\\
					1818181.81818182	7.65500714612921e-05\\
					1848484.84848485	0.000105974935230072\\
					1878787.87878788	0.00014666895016661\\
					1909090.90909091	0.000202910100931952\\
					1939393.93939394	0.000280565877643019\\
					1969696.96969697	0.000387652565197921\\
					2000000	0.000535062962039357\\
					2030303.03030303	0.00073748715211483\\
					2060606.06060606	0.00101452815921586\\
					2090909.09090909	0.00139196023902728\\
					2121212.12121212	0.0019029696732842\\
					2151515.15151515	0.00258902825112224\\
					2181818.18181818	0.00349975835775307\\
					2212121.21212121	0.00469078222463624\\
					2242424.24242424	0.00621825376101104\\
					2272727.27272727	0.00812890934912845\\
					2303030.3030303	0.01044561083506\\
					2333333.33333333	0.013150914456988\\
					2363636.36363636	0.0161746116460833\\
					2393939.39393939	0.0193930990956707\\
					2424242.42424242	0.0226456579149648\\
					2454545.45454545	0.0257647331363239\\
					2484848.48484848	0.0286089672675848\\
					2515151.51515152	0.0310859693149977\\
					2545454.54545455	0.0331581963719954\\
					2575757.57575758	0.0348343056898979\\
					2606060.60606061	0.036153422053872\\
					2636363.63636364	0.0371693987360435\\
					2666666.66666667	0.0379389578423718\\
					2696969.6969697	0.0385145347439157\\
					2727272.72727273	0.0389409639215144\\
					2757575.75757576	0.0392546793998384\\
					2787878.78787879	0.0394842814402344\\
					2818181.81818182	0.0396516863633951\\
					2848484.84848485	0.0397734055441371\\
					2878787.87878788	0.0398617289372403\\
					2909090.90909091	0.0399257257050585\\
					2939393.93939394	0.0399720469851935\\
					2969696.96969697	0.040005548945783\\
					3000000	0.0400297658787313\\
				};
				\addlegendentry{$T = \SI{60}{\degreeCelsius}$}
				
			\end{axis}
		\end{tikzpicture}
		\caption{Field- and temperature-dependent conductivity $\sigma(E,T)$ of the nonlinear material. The electric field strength and the temperature are denoted as $E$ and $T$, respectively.}
		\label{fig:sigma_general}
	\end{figure}
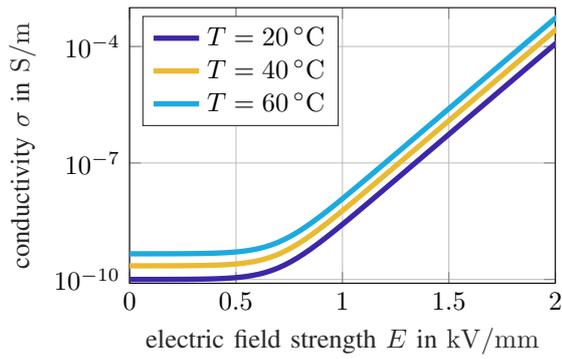
	
	\begin{figure}
		\setlength{\fwidth}{0.6 \linewidth}
		\definecolor{mycolor1}{rgb}{0.24220,0.15040,0.66030}%
		\begin{tikzpicture}
			\begin{axis}[
				width=1.15\fwidth,
				height=0.75\fwidth,
				at={(0\fwidth,0\fwidth)},
				scale only axis,
				ytick={150,450,750},
				xlabel style={font=\color{white!15!black}},
				yminorticks=true,
				axis background/.style={fill=white},
				xmajorgrids,
				ymajorgrids,
				yminorgrids,
				xlabel={time $t$ in \unit{\micro\second}},
				ylabel={voltage $V(t)$ in \unit{\kilo\volt}}]
				\addplot+[color=mycolor1, line width=2.0pt, mark = none] table[x=time, y=value, y expr = \thisrowno{1} + 150] {impulse_data.dat};
			\end{axis}
		\end{tikzpicture}
		\caption{Transient voltage $V(t)$ over the time $t$.}
		\label{fig:impulse}
	\end{figure}
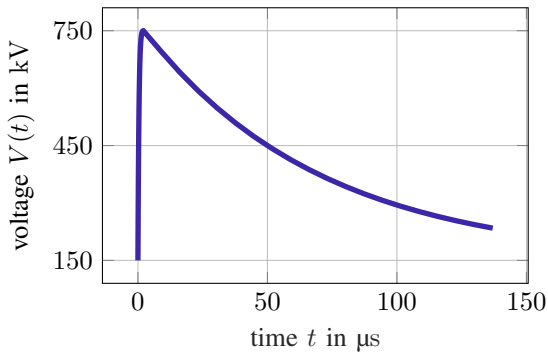
	
	\begin{figure}
		\centering
		\includegraphics[width=0.78\linewidth]{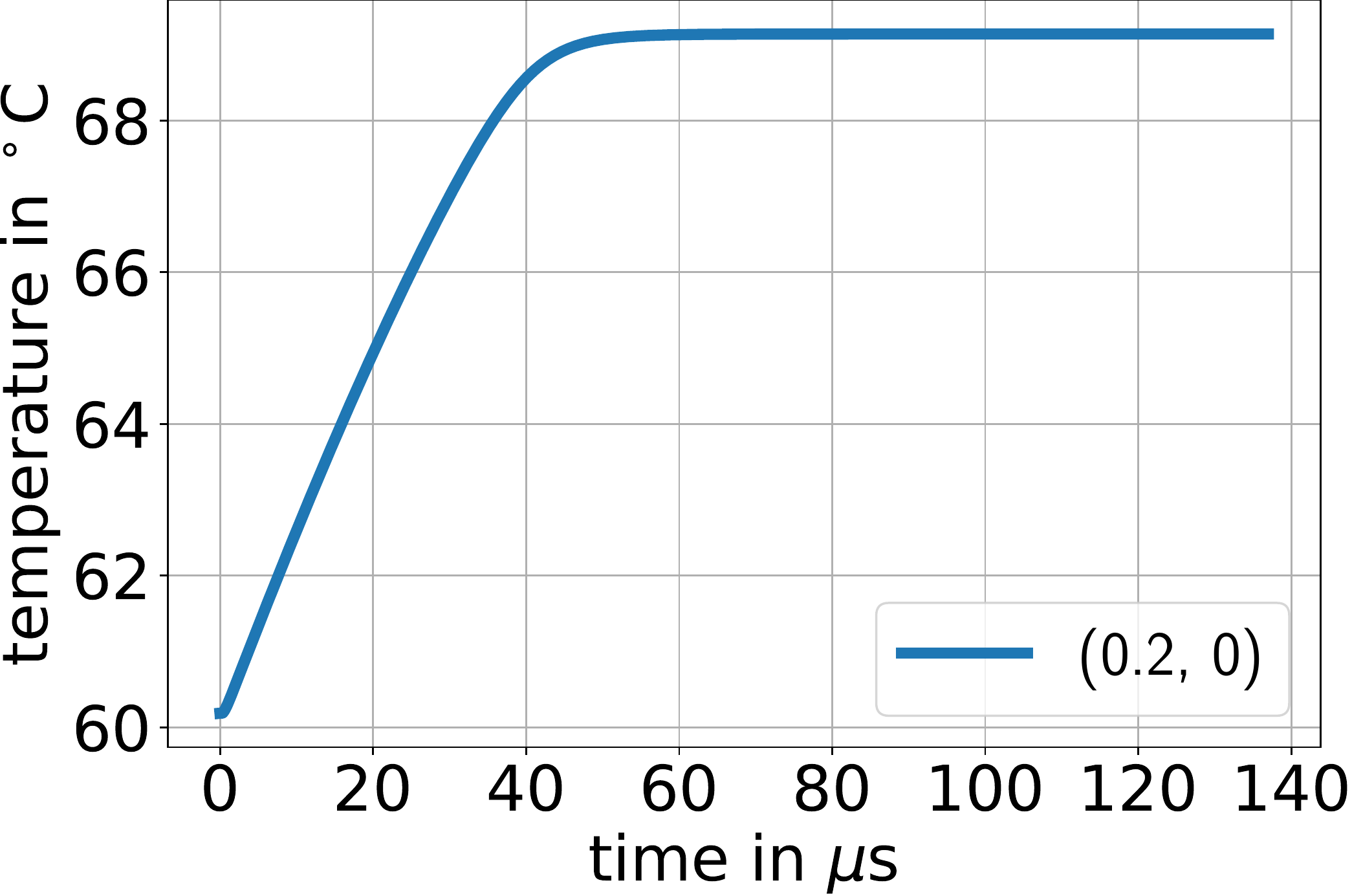}
		\caption{Plot generated by the method \function{plot\_\-temperature\_\-at\_\-position} of the temperature evaluated at $\rho = \SI{0.2}{\meter}$ inside the resistor depicted in Fig.~\ref{fig:ex1_setup} over time.}
		\label{fig:ex1_plot}
	\end{figure}
	
	The second example is a nonlinear electrothermal simulation of a cylindrical resistor (see Fig.~\ref{fig:ex1_setup}). The resistor consists of a nonlinear material, which features a strongly field- and temperature-dependent electric conductivity (see Fig.~\ref{fig:sigma_general}). Such nonlinear materials are employed in field grading layers of high-voltage cable systems \parencite{Hussain_2017aa, Ruppert_2023ab}. The material is located between two electrodes to which a transient voltage $V(t)$ is applied (see Fig.~\ref{fig:impulse}). The resistor is surrounded by a layer of soil. %
	The inner surface of the resistor and the outer surface of the soil are at a fixed temperature of \SI{60}{\degreeCelsius} and \SI{20}{\degreeCelsius}, respectively. %
	For a more detailed description of the model configuration and material parameters see \parencite{Ruppert_2023ab}.
	
	The first step in the simulation process is the definition of the geometry. \pyrit offers several ways to do this: The user can import a step file, create a geometry using \pyrit's \module{geometry} package, or import an already existing \gmsh file. For this example, an already existing \gmsh file is imported:
	\begin{pythoncode}
		# Import mesh from GMSH file
		filename = './gmsh/cylindrical_resistor'
		mesh = read_msh_file(filename, mesh_type=AxiMesh)
	\end{pythoncode}
	
	Second, the materials and boundary conditions are defined. The excitation voltage and fixed temperatures are implemented as Dirichlet boundary conditions of the electric potential and temperature, respectively:
	\begin{pythoncode}
		# Define materials
		field_grading_material = MaterialLibrary.get_material("FGM")
		soil = Mat("Soil", ThermalConductivity(0.8), VolumetricHeatCapacity(2333e3))
		materials = Materials(field_grading_material, soil)
		
		# Define boundary conditions of the electric subproblem
		ground = BCDirichlet(0, "ground")
		voltage = BCDirichlet( ExcitationLibrary.get_excitation( "lightning_impulse"), "V")
		boundary_conditions_electric = BdryCond(ground, voltage)
		
		# Define boundary conditions of the thermal subproblem
		temperature_fgm = BCDirichlet(333.15, "temperature_inner")
		temperature_soil = BCDirichlet(293.15, "temperature_outer")
		boundary_conditions_thermal = BdryCond(temperature_fgm, temperature_soil)
	\end{pythoncode}
	The homogeneous Neumann boundary conditions of the lower and upper boundary in the $z$-direction are automatically satisfied by the \ac{fe} ansatz.
	
	Finally, the material properties and boundary conditions are assigned to the geometry. A region maps the physical properties to corresponding edges and elements in the \ac{fe} mesh: 
	\begin{pythoncode}
		# Define regions of the electric subproblem
		regions_electric = generate_regions(filename, materials, boundary_conditions_electric)
		
		# Define regions of the thermal subproblem
		regions_thermal = generate_regions(filename, materials, boundary_conditions_thermal)
	\end{pythoncode}
	
	As stated in Sec.~\ref{sec:software_structure}, \pyrit offers classes for \ac{em} and heat transfer problems of any time de\-pen\-dence. This allows to solve field problems based on a template-based workflow. Here, the class \class{Electrothermal\-Problem\-Axi\-Transient}, that represents a coupled transient electrothermal problem in cylindrical coordinates, is utilized. It is composed of two subproblems containing the boundary conditions of the electric and thermal subsystem, respectively:
	\begin{pythoncode}
		# Define time axis
		time_steps = numpy.append(numpy.linspace(0,2e-6,150), numpy.linspace(2e-6, 140e-6, 150)[1:])
		
		# Define the electro-quasistatic subproblem
		problem_electric = CurrentFlowProblemAxiTransient( "cylindrical_resistor", mesh, regions_electric, materials, boundary_conditions_electric, time_steps)
		
		# Define the transient heat conduction subproblem
		problem_thermal = ThermalProblemAxiTransient( "cylindrical_resistor", mesh, regions_thermal, materials, boundary_conditions_thermal, time_steps)
		
		# Combine the subproblems to a coupled electrothermal problem
		problem_coupled = ElectrothermalProblemAxiTransient( problem_electric, problem_thermal)
	\end{pythoncode}
	
	The \class{ElectrothermalProblemAxiTransient} class offers a \function{solve} method for running the electrothermal simulation. It automatically handles the coupling of the two subproblems either by a weak coupling scheme or a successive substitution iteration. The nonlinearity arising in the electric subproblem due to the nonlinearity of the nonlinear material is solved using a damped Newton method:
	\begin{pythoncode}
		# Load previously computed steady-state solutions that serve as initial conditions for the electric potential and temperature
		initial_value_potential = Solution.load("steady_state_electric.pkl")
		initial_value_temperature = Solution.load("steady_state_thermal.pkl")
		
		# Run simulation
		solution_electric, solution_thermal = problem_coupled.solve( initial_value_potential, initial_value_temperature)
	\end{pythoncode}

	The solutions returned by the \function{solve} method come with a set of post-processing and plotting routines:
	\begin{pythoncode}
		# Print power loss during peak excitation
		print(f"Power loss: {solution_electric.joule_loss_power( time=2e-6)} W")
		# Animate the electric potential over time
		solution_electric.animate_potential()
		# Plot the temperature at a specified position over time
		position = (0.2, 0)
		solution_thermal.plot_temperature_at_position( position, celsius=True)
	\end{pythoncode}
	
	Figure~\ref{fig:ex1_plot} exemplarily shows the plot generated by the method \function{plot\_temperature\_at\_position}.
	
	\FloatBarrier
	\subsection{Customized example} %
	
	\setcounter{FancyVerbLine}{0}%
	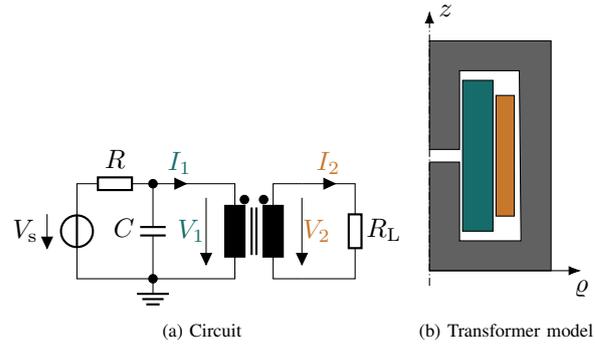
\begin{figure}[t]
		\def\primcolor{cyan4}%
		\def\seccolor{brown6}
		\def\spycolor{red}
		\def\yokecolor{gray4}
		\newcommand{\Ione}{\textcolor{\primcolor}{I_1}}%
		\newcommand{\Vone}{\textcolor{\primcolor}{V_1}}%
		\newcommand{\Itwo}{\textcolor{\seccolor}{I_2}}%
		\newcommand{\Vtwo}{\textcolor{\seccolor}{V_2}}%
		\centering
		\subcaptionbox{Circuit\label{fig:circuit}}{
			\ctikzset{resistors/scale=0.4,inductors/scale=0.6,sources/scale=0.5,capacitors/scale=0.4}%
			\begin{circuitikz}
				\def\disttotrafo{0.7}
				\def\tmpdist{1}
				\draw (0,0) node[transformer core] (trafo) {} (trafo.inner dot A1) node[circ]{} (trafo.inner dot B1) node[circ]{};
				\draw (trafo.A1) to[short,i<_={$\Ione$}] ++(-\disttotrafo,0) node[circ] (a1o) {};
				\draw (trafo.A2) -- ++(-\disttotrafo,0) node[circ] (a2o) {} -- ++(0,-0.2) node[tlground] {};
				\node[draw, ground] at (a2o) {};
				
				\draw (trafo.A1) to[open, v={$\Vone$},straight voltages,voltage shift=-1.0] (trafo.A2);%
				\draw (trafo.B1) to[open, v^={$\Vtwo$},straight voltages,voltage shift=-1.0] (trafo.B2);
				
				\draw (a1o) to[capacitor,l_=$C$] (a2o);
				\draw (a1o) to[european resistor,l_=$R$] ++(-\tmpdist,0) node (tmpa) {} to[european voltage source, v_={$V_\mathrm{s}$},voltage shift=1.0] (tmpa |- a2o) -- (a2o);
				
				\draw (trafo.B1) to[short,i>^={$\Itwo$}] ++(\disttotrafo,0) node (tmpb){} to[european resistor,l=$R_\mathrm{L}$] (tmpb |- trafo.B2) -- (trafo.B2);
				
				\def\tmplen{0.15}
				\def\tmpheight{0.05}
			\end{circuitikz}%
		}
		\subcaptionbox{Transformer model\label{fig:trafo_model}}{
			\begin{tikzpicture}[x=0.04cm,y=0.04cm,]
				\def\width{40}
				\def\height{76.2}
				\def\widthinner{9.9}
				\def\widthbot{9.9}
				\def\widthtop{9.9}
				\def\widthouter{10.45}
				\def\airgap{4.2}
				\def\widthfoil{10}
				\def\heightfoil{50}
				\def\widthwire{6}
				\def\heightwire{40}
				\def\foiltoinner{1}
				\def\wiretoinner{12}
				
				\pgfmathparse{\widthinner + \foiltoinner}
				\pgfmathsetmacro{\foilbrx}{\pgfmathresult}
				\pgfmathparse{\height/2-\heightfoil/2}
				\pgfmathsetmacro{\foilbry}{\pgfmathresult}
				
				\pgfmathparse{\widthinner + \wiretoinner}
				\pgfmathsetmacro{\wirebrx}{\pgfmathresult}
				\pgfmathparse{\height/2-\heightwire/2}
				\pgfmathsetmacro{\wirebry}{\pgfmathresult}
				
				\draw[fill=gray4] (0,0) -- (\width,0) -- (\width,\height) -- (0,\height) -- (0,0.5*\height+0.5*\airgap) -- ++(\widthinner,0) -- (\widthinner,\height-\widthtop) -- (\width-\widthouter,\height-\widthtop) -- (\width-\widthinner,\widthbot) -- (\widthinner,\widthbot) -- (\widthinner,0.5*\height-0.5*\airgap) -- ++(-\widthinner,0) -- (0,0);
				\draw[fill=\primcolor] (\foilbrx,\foilbry) rectangle ++(\widthfoil,\heightfoil);
				\draw[fill=\seccolor] (\wirebrx,\wirebry) rectangle ++(\widthwire,\heightwire);
				
				\def\distaxis{10}
				\draw[->] (0,0) -- ++(\width + \distaxis,0) node[below] {$\varrho$};
				\draw[densely dash dot,->] (0,-0.5*\distaxis) -- (0,\height+\distaxis) node[right] {$z$};
				
				\def\tmpsep{2}
				
			\end{tikzpicture}%
		}
		\caption{Configuration of the third example. The circuit for the field-circuit coupling (\subref{fig:circuit}) with a transformer resolved in a field model (\subref{fig:trafo_model}). The transformer is described in axisymmetric coordinates and consists of a highly permeable yoke (gray) and a primary windings (teal) and secondary windings (orange).}
		\label{fig:trafo_example}
	\end{figure}
	In the third example, no predefined problem class is used.
	A field-circuit coupled problem with a magnetoquasistatic field problem is considered (Fig.~\ref{fig:trafo_example}).
	It consists of an electrical circuit (see Fig.~\ref{fig:circuit}) with a transformer that is represented by a field model (see Fig.~\ref{fig:trafo_model}). 
	The lumped elements in the circuit are $R=\SI{1}{\ohm}$, $R_\mathrm{L}=\SI{10}{\ohm}$ and $C=\SI{0.1}{\micro\farad}$. The voltage source is $V_\mathrm{s}=\hat{V}_\mathrm{s}\sin\left(\omega t\right)$, with $\hat{V}_\mathrm{s}=\SI{1}{\volt}$, the angular frequency $\omega=2\pi f$, the frequency $f=\SI{3}{\kilo\hertz}$ and the time variable $t$. 
	The field model of the transformer uses a stranded conductor model for both coils \parencite{Gersem2004ab}. The primary coil (teal) and the secondary coil (orange) have \num{300} and \num{500} turns, respectively.
	
	In the following, one system for the magnetoquasistatics and one for the circuit is built in the frequency domain. 
	These are then coupled with field-circuit coupling matrices.
	
	The geometry and the mesh of this example were generated and exported to \pyrit with the \gmsh Python API.
	Since the focus of this example is on the field-circuit coupling, these steps are not shown here. We start with a \pythoninline|problem| object on which the shape functions, the mesh and  the materials and boundary conditions already exist:
	\begin{pythoncode}
		problem = model.create_geometry()
	\end{pythoncode}
	
	Now, the right-hand side vector and the \ac{fe} matrices are retrieved from the shape function object and combined into the system of equations of the magnetoquasistatic formulation:
	\begin{pythoncode}
		curlcurl = shape_function.curlcurl_operator(regions, materials, Reluctivity)
		mass = shape_function.mass_matrix(regions, materials, Conductivity)
		rhs = shape_function.load_vector(0)
		
		matrix = curlcurl + 1j * omega * mass
	\end{pythoncode}
	
	The system representing the circuit (see Fig.~\ref{fig:circuit}) is built with the values of the lumped elements:
	\begin{pythoncode}
		circuit_matrix = sparse.coo_matrix(
		numpy.array([[R,1+R*1j*omega*C,0   ,0],
		[0,0             ,-R_L,1]]))
		circuit_rhs = sparse.coo_matrix(([V_s,], ([0,], [0,])), shape=(2, 1))
	\end{pythoncode}
	
	The coupling matrices from the field-circuit coupling are generated by a class that represents stranded conductors:
	\begin{pythoncode}
		bottom_matrix, right_matrix, \
		diagonal_matrix, rhs_vector = 
		str_cond.get_coupling_matrices(problem)
	\end{pythoncode}
	
	Inserting the boundary conditions with the method \pythoninline|shrink| yields the system matrix and the system vector. With these, the new system is given by:
	\begin{pythoncode}
		matrix, rhs = problem.shape_function.shrink(matrix, rhs, problem)
		
		matrix_sys = sparse.bmat(
		[[matrix_shrink, right_matrix],
		[bottom_matrix, diagonal_matrix]])
		rhs_sys = sparse.vstack([rhs_shrink, rhs_vector])
	\end{pythoncode}
	
	Finally, we insert the circuit system:
	\begin{pythoncode}
		matrix_sys = sparse.vstack(
		[matrix_sys, 
		sparse.hstack([zeros, circuit_matrix])])
		rhs_sys = sparse.vstack(
		[rhs_sys, circuit_rhs])
	\end{pythoncode}
	
	After solving the system, the circuit solution consists of the voltages and currents at the transformer. The boundary conditions are integrated into the field solution that is then evaluated with a predefined solution class for magnetoquasistatic problems:
	\begin{pythoncode}
		sol = problem.solve_linear_system( matrix_sys, rhs_sys)[0]
		
		I_1, V_1, I_2, V_2 = sol[-4:]
		solution = inflate(sol, problem)
	\end{pythoncode}
	
	The \pythoninline|solution| object provides post-processing routines:
	\begin{pythoncode}
		print(f"Energy: {solution.energy()} J")
		solution.plot_b_field('abs', mesh=False)
	\end{pythoncode}
	
\begin{figure}
		\centering
		\includegraphics[scale=1]{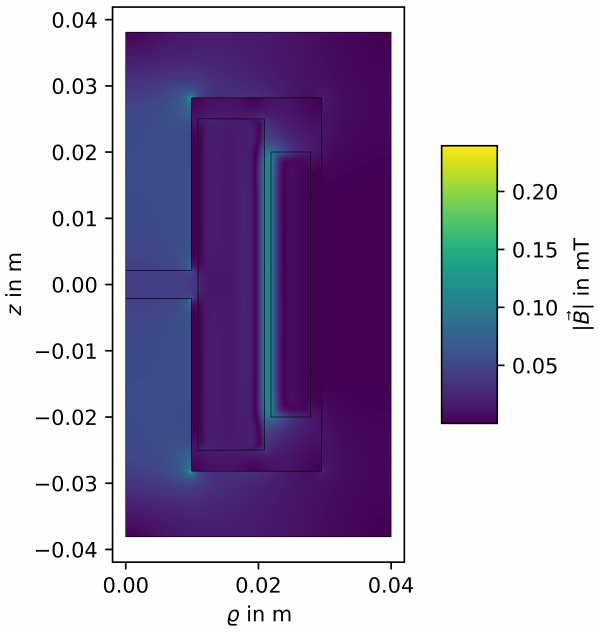}
		\caption{Absolute values of the magnetic flux density at phase 0 in the transformer.}
		\label{fig:trafo_plot}
	\end{figure}

	The plot of the magnetic flux density at a phase of 0, generated by the method \pythoninline|plot_b_field|, is shown in Fig.~\ref{fig:trafo_plot}.
	
	\section{Conclusion}
	\pyrit is a \acl{fe} software environment for simulating \acl{em} and thermal field problems of any time dependence (static, transient or harmonic).
	Its software structure is based on an engineering workflow for solving a field problem numerically, i.e., problem definition, problem solution and post-processing. \pyrit allows to solve field problems in a template-based workflow. 
	Furthermore, full access to  the \ac{fe} core supports research and development on numerical techniques.
	Three examples showed how users can define, solve and post-process a field problem in an easy and general way.
	
	\section{Acknowledgment}
	This work is supported by the joint DFG/FWF Collaborative Research Centre CREATOR (CRC -- TRR361/F90) at TU Darmstadt, TU Graz and JKU Linz, the Athene Young Investigator Programme of TU Darmstadt, the German Science Foundation (DFG project 436819664), and the Graduate School Computational Engineering at TU Darmstadt. 
	The authors thank Christian Bergfried, Daniel Leißner, Jonas Christ and Svenja Menzenbach for their valuable contributions. In particular, the authors thank Laura A. M.\ D'Angelo, who provided her knowledge from developing the Matlab \ac{fe} tool \textit{Niobe} and Herbert De Gersem for the opportunity to pursue this project and for his valuable advise. Furthermore, the authors thank Michael Leigsnering for fruitful discussions on collaborative software development and for the idea of the software's name.
	
	\printbibliography

	\begin{acronym}
		\acro{2d}[2D]{two-dimensional}
		\acro{3d}[3D]{three-dimensional}
		\acro{ac}[AC]{alternating current}
		\acro{dc}[DC]{direct current}
		\acro{dof}[DoF]{degree of freedom}
		\acroplural{dof}[DoFs]{degrees of freedom}
		\acro{em}[EM]{electromagnetic}
		\acro{epdm}[EPDM]{ethylene propylene diene monomer rubber}
		\acro{eqs}[EQS]{electroquasistatic}
		\acro{eqst}[EQST]{electroquasistatic-thermal}
		\acro{es}[ES]{electrostatic}
		\acro{fe}[FE]{finite-element}
		\acro{fem}[FEM]{finite-element method}
		\acro{fgm}[FGM]{field grading material}
		\acro{hv}[HV]{high-voltage}
		\acro{hvac}[HVAC]{high-voltage alternating current}
		\acro{hvdc}[HVDC]{high-voltage direct current}
		\acro{lsr}[LSR]{liquid silicone rubber}
		\acro{sir}[SiR]{silicone rubber}
		\acro{mo}[MO]{metal oxide}
		\acro{pd}[PD]{partial discharges}
		\acro{pde}[PDE]{partial differential equation}
		\acro{pea}[PEA]{pulsed-electro-acoustic method}
		\acroplural{pde}[PDEs]{partial differential equations}
		\acro{pm}[PM]{person month}
		\acroplural{pm}[PMs]{person months}
		\acro{qoi}[QoI]{quantity of interest}
		\acroplural{qoi}[QoIs]{quantities of interest}
		\acro{rdm}[RDM]{research data management}
		\acro{rms}[rms]{root mean square}
		\acro{tna}[TNA]{Transient Network Analysis}
		\acro{tuda}[TUDa]{TU~Darmstadt}
		\acro{tum}[TUM]{TU~Mün\-chen}
		\acro{uq}[UQ]{uncertainty quantification}
		\acro{wp}[WP]{work package}
		\acroplural{wp}[WPs]{work packages}
		\acro{xlpe}[XLPE]{cross-linked polyethylene}
		\acro{zno}[ZnO]{zinc oxide}
		\acro{CI}[CI]{continuous integration}
	\end{acronym}
	
\end{document}